\documentclass[11pt,a4paper]{article}
\usepackage{jheppubnohead}
\usepackage{graphicx}
\usepackage{bm}

\preprint{APS/123-QED}

\title{Gravitational to Coulomb force ratio and the origin of the Cosmic magnetic field}

\author{N. D. Padilla$^{1,2}$}
\emailAdd{nelson.padilla@unc.edu.ar}
\author{J. Racker$^{1,2}$}
\author{I. J. Araya$^{3}$}
\author{F. Stasyszyn$^{1,2}$}
\affiliation{$^1$ Instituto de Astronomía Teórica y Experimental, IATE, CONICET-UNC, Laprida 854, X5000BGR, Córdoba, Argentina\\
$^2$Observatorio Astronómico de la Universidad Nacional de Córdoba, Laprida 854, X5000BGR, Córdoba, Argentina\\
$^3$ Universidad Andres Bello, Departamento de Física y Astronomía, Facultad de Ciencias
Exactas, Sazié 2212, Piso 7, Santiago, Chile\\
}

\abstract{The origin of the seeds of galactic magnetic fields is a subject that remains under debate. Here we will explore a{ simple source} based on tiny charge asymmetries in slowly rotating protogalaxies. We use current knowledge of galaxy formation and evolution to estimate that a charge imbalance of $1$ every $\sim 10^{38\pm 5}$ charge carriers in slowly rotating protogalaxies can provide adequate seeds for the galactic dynamos. Interestingly, this is of the same order than the ratio of gravitational to Coulomb forces between the elementary plasma
constituents. Motivated by this fact, we study different mechanisms for generating such charge imbalances from a {direct} interplay of gravitational and Coulomb forces, namely the possibility that these are of primordial origin, that stellar or primordial black holes redistribute charge in protogalaxies, or that the imbalance is sourced by gravity as the galaxy forms in quasi-hydrostatic equilibrium. Our results show that primordial asymmetries drop to small values by the onset of galaxy formation,  with an amplitude that is similar to the possible charge asymmetries that could be produced by black holes.  Although these charge asymmetries can have values within the range of interest, they are much smaller than the gravitationally induced one in hydrostatic equilibrium conditions. The latter lies in the upper range of the required charge imbalance.} 
\begin{document}

\maketitle


\section{Introduction}
Magnetic fields are ubiquitous in the Universe 
with present-day amplitudes that can be explained by astrophysical dynamos driven by the baryonic plasma processes that take place after galaxy formation begins (see e.g.~the reviews \cite{Kulsrud99, Brandenburg05, Brandenburg2023}).  

However, these fields require a seed since amplification processes require a non-zero initial field. These seeds have been proposed to be produced by several mechanisms and are generally divided into primordial and late-time seeds. The former correspond to mechanisms that could take place early after the Big Bang \cite{Grasso01,Widrow02,medvedev}, or during phase transitions \cite{Quashnock89,Vachaspati91}, due to early non-linearities in the pre-decoupling  primordial density perturbations \cite{Matarrese2005,gopal2005,takahashi2005,Saga15}, or even primordial black holes \cite{araya}.  There are also proposals for charge asymmetries involving non-Standard Model particles \cite{DolgovSilk} or arising from massive photons and black hole evaporation \cite{Dolgov} that could produce the seed magnetic fields. On the other hand, late-time seeds include plasma \cite{zhou} and Biermann battery processes \cite{Biermann:1950} occurring, for instance, due to shocks during gravitational collapse \cite{Subramanian1994, Davies2000, Kulsrud08, Attia21} and due to star formation including population III stars \cite{xu,Durrive2015}, and those induced by the extraction of charge by black holes of primordial or stellar origin from the protogalactic plasma~\cite{Padilla24}. Further details and references can be found in reviews such as \cite{Widrow02, carilli02, Govoni04, vallee04, Barrow07, Widrow11, Durrer13, Subramanian16, Vachaspati21} ({see also the discussion section}).  

One possible way to tell apart different seeding scenarios is to compare their amplitudes and scales of influence with those required to explain the present-day magnetic field configurations. There have been several attempts at estimating the level of enhancement of magnetic fields due to the dynamo processes in order to estimate the required seed fields. In~\cite{Davis1999} an estimate was made allowing for the first time for the effect of a cosmological constant, which effectively increases the age of the Universe and, consequently, allowed to consider very small seed values.  They found that the level of required seed is of $B_{\rm seed}\sim10^{-30\pm5}$~G \cite{Davis1999}, where the uncertainty is mostly due to the difficulty in assessing the e-folding timescale of the astrophysical dynamo enhancement of the field. Prior estimates were quite larger than this value.  The seed proposals listed above lie within and above this minimum seed estimate.

In this work we will study the possibility that a magnetic seed arises from charge asymmetries in  protogalaxies  which are seen to rotate, possibly due to the effect of tidal fields from neighboring overdensities prior to their collapse \cite{peebles} {combined with dissipative effects.} 
We will first calculate, in Sec.~\ref{sec:seedlevel}, the required level of charge asymmetry in rotating protogalaxies in order to produce adequate magnetic field seeds. This calculation shows an interesting match with the gravitational to Coulomb force ratio between the plasma species. Therefore, in Sec.~\ref{sec:origin} we analyze different ways for generating charge imbalances which involve a simple interplay of
gravitational and Coulomb forces, {so that the size of the resultant charge asymmetry can be clearly related to the ratio between gravitational and Coulomb forces}.
We first {consider possible primordial charge asymmetries, evaluating the largest amount} that could survive until the epoch of galaxy formation, then we {study} asymmetries driven by black holes within galaxy dark matter haloes, and finally {we analyze} the asymmetry due to hydrostatic equilibrium.
We will show that the -small- charge asymmetry that must be present in hydrostatic equilibrium~\cite{Pannekoek22, Rosseland24, Eddington1926} (see also, e.g., \cite{Belmont2013,Parks2018}), 
is significantly larger than the other two possibilities explored here, 
and that it lies within the upper range required to seed magnetic fields of the size calculated in \cite{Davis1999} (while still being small enough to allow the hot plasma in galaxies to be considered neutral). Then, in Sec.~\ref{sec:discussion}, we discuss the previous results and compare with other proposals for the origin of the seeds of galactic magnetic fields. Finally, we conclude in Sec.~\ref{sec:conclusions}.

\section{Required level of charge asymmetry}

\label{sec:seedlevel}
To estimate the amplitude of a seed field arising from a charge asymmetry in a slowly rotating primeval galaxy, we make a calculation of the magnetic field resulting from the rotation of the plasma bound to the protogalaxy that collapsed along with the dark matter that is thought to dominate gravitationally bound objects in the Universe~\cite{Rubin1978,Planck:2018vyg}.
The following  assumes that gas, 
and therefore, baryons, are mostly in a hot gas state.  We will model the charge and spin of the hot gas phase but the effective magnetic field that we will concentrate on is the one present  soon after gas cooling proceeds and a primeval disc forms.  Dynamo processes would start to take place within this disc, which justifies concentrating on magnetic fields directly affecting it.  
Even though discs typically present angular rotation velocities much higher than that of the hot gas halo, we only look at the rotation of the latter as this is the fluid that is proposed to contain a charge asymmetry here.  
We adopt the following assumptions, which reflect the current model of galaxy formation and of the structure of dark matter dominated objects in the Universe.

We assume that the hot baryonic plasma follows a core profile  \cite{Lacey16} forming a singular isothermal sphere \cite{springel}.   If the magnetic moment of the plasma is sourced by a charge asymmetry 
\begin{equation}
    \epsilon_{q}=n_i/n_j-1,
\end{equation} where $n$ represents a number density and $i$ and $j$ correspond to either electrons or protons, depending on the charge sign of the asymmetry, the resulting magnetic moment amplitude reads
\begin{equation}
\label{eq:magmom}
    m(r)=\left|\frac{1}{2}\int_V \vec r \times (\rho_{q} \vec v) dV \right|,
\end{equation}
where  the charge density is,
\begin{equation}
    \rho_{q}=\epsilon_{q} e \rho_b/(\mu m_p).
    \label{eq:rhoq}
\end{equation} 
Here $e$ is the  proton charge, $m_p$ is the proton mass, $\mu\simeq1.4$ the mean molecular weight of the plasma, and $\rho_b$ the baryon mass density.

Moreover, we assume that the hot plasma has the same specific angular momentum than the dark matter \cite{maccio07}, which has a dimensionless spin parameter well approximated by $\lambda'=J_{\rm vir}/(\sqrt2 M_{\rm vir}V_{\rm vir}r_{\rm vir})$, where $M_{\rm vir}$, $J_{\rm vir}$, $V_{\rm vir}$ and $r_{\rm vir}$ are the mass, angular momentum, virial velocity and virial radius of the dark matter halo host, respectively.  These approximations are commonly used in models of galaxy formation and evolution which are able to reproduce the sizes and spins of galaxies \cite{Baugh2006}.  Notice that the rotation of galaxies is quite faster than that of this primeval hot gas atmosphere which has not yet contracted due to cooling.  

This results in the following amplitude of a dipolar magnetic field for the plasma at the virial radius of the galaxy, representative of the extent of the {primeval} hot gas halo,
\begin{equation}
    B(r_{\rm vir})=\frac{\mu_0}{4\pi}\frac{m(r_{\rm vir})}{r_{\rm vir}^3}=\frac{\Omega_b}{\Omega_m}\frac{\mu_0}{4\pi}\frac{1}{2}\frac{\epsilon_{q} e}{\mu m_p} \lambda' \sqrt{2}\frac{M_{\rm vir}V_{\rm vir}}{r_{\rm vir}^2},\label{eq:B}
\end{equation}
 where  $\mu_0$ is the magnetic permeability of the vacuum, and the baryon fraction represented by the ratio of the baryon to matter density parameters, $\Omega_b/\Omega_m$, multiplying the virial mass, $M_{\rm vir}$, is used as an estimate of the mass of the plasma in the protogalaxy, ignoring cooling and other processes that become important with the onset of star formation.  The virial radius can be obtained using the virial mass and the typical halo overdensity (see for instance \cite{Bryan_1998}),  and the virial velocity follows from the virial relations.

Next we take a dimensionless galaxy spin parameter $\log_{10} \lambda'=-1.5$ following \cite{maccio07}, and adopt the history of mass accretion from \cite{Fakhouri10} for a Milky Way-like halo to estimate the Galaxy mass at an approximate redshift for its initial collapse of $z\sim 20$, which results in $M_{\rm vir}\sim 3 \times 10^{9}M_\odot$. 

{Plugging all this into Eq.~\eqref{eq:B}, taking for $B(r_{\rm vir})$
the range of values which according to~\cite{Davis1999} can seed galactic dynamos, and solving for $\epsilon_q$, we obtain~\footnote{Actually, the magnetic field at the disc, i.e.~approximately at the center of the hot gas halo, is roughly a factor 5 larger than the dipolar magnetic field $B(r_{\rm vir})$ that we have taken as representative for the system.}}
\begin{equation}
    \epsilon_{q}\sim 10^{-38\pm5}.\label{Eq:req}
\end{equation}

Notice that the ratio of gravitational to Coulomb forces between the elementary plasma constituents lies in this range, e.g. that between two protons is  
\begin{equation}
    F_G/F_C=\frac{G\, m_p^2}{k_C\, e^2}\sim 10^{-36}, \label{Eq:CG}
\end{equation}
where  $G$ and $k_C$ are the gravitational and Coulomb constants, respectively.  This number represents a remarkable coincidence with the required charge imbalance for the magnetic seed.

We now turn to the possible origins of charge asymmetries $\epsilon_q$ in protogalaxies discussing, firstly, upper limits on the amplitude of primordial spatial charge fluctuations by the epoch of galaxy formation (these limits do not apply to global charge asymmetries), secondly, {asymmetries from charge displacement by black holes in protogalaxies and, thirdly, the charge imbalance arising from hydrostatic equilibrium.} 

\section{Origin of charge asymmetries}
\label{sec:origin}
\subsection{Primordial charge asymmetries}
\label{sec:primordial}
Different mechanisms for primordial (local and non-local) charge fluctuations have been proposed, see e.g.~\cite{Siegel2,DolgovSilk,Lozanov,Donofrio,Soriano}, which are consistent with the observational bounds on a net electric charge asymmetry of the Universe~\cite{Orito:1985cf,Caprini:2003gz,Masso:2002vh}. Although these primordial fluctuations could potentially provide 
a charge imbalance within protogalaxies, it is not clear whether they would be able to survive until the epoch of galaxy formation.
Instead of exploring the possible amplitudes for these primordial spatial charge fluctuations, we directly estimate the maximum charge imbalance within an expanding background leftover after decoupling, that would eventually reach turnaround, using the Newtonian approximation for the estimate of the critical density of the Universe, adapted to Coulomb forces for the unbalanced charges and assuming that electrons, the lighter charge carriers, would follow the positive charge carriers.  If turnaround is attained we assume the charge imbalance would be neutralized.

Given that most of the matter is in a neutral state since decoupling down to the epoch of reionization, we assume that the unbalanced charges are fully decoupled from the cosmic background radiation and that they respond only to their mutual Coulomb forces and to the total gravitational potential, ignoring for simplicity other possible interactions (see e.g. \cite{Dolgov}). At the redshifts of interest here, these considerations result in 
\begin{equation}
    \frac{\dot a^2}{a^2}     = \frac{8 \pi}{3} \left(G \rho + \frac{k_C \mu m_p \rho_{q}^2}{m_e\rho_b \epsilon_{q}}\right) \nonumber \\
                             = H_{\Lambda CDM}(z)+\frac{8 \pi }3 \frac{k_C \mu m_p \rho_{q}^2}{m_e \rho_b \epsilon_{q}},
\end{equation}
where $\rho$ is the total energy density and $a$ represents the scale factor of the region that circumscribes the unbalanced charges with density $\rho_q$. In the second equality we replaced the first term with the $\Lambda$CDM Hubble factor. Equating the two terms in the right hand side of this equation, we obtain the maximum imbalance that would be able to expand along with the global $\Lambda$CDM expansion, regardless of the sign of the imbalance, i.e.,
\begin{equation}
\epsilon_{q}^{coll}=\frac{3 \mu m_p m_e H_{\Lambda CDM}(z)^2}{8 \pi k_C \rho_{b,0} (1+z)^3 e^2}=\frac{G\mu m_pm_e}{k_C e^2}\frac{\rho_m}{\rho_b},
\label{eq:epsdelta}
\end{equation}
where in the second equality we have assumed the total energy density to be comparable to that of the mass, $\rho\simeq\rho_m$, and Eq. \ref{eq:epsdelta} shows that this excess is within an order of magnitude of the gravitational to Coulomb force ratio between electrons and protons, i.e.~it is about three orders of magnitude smaller than the 
{charge imbalance to be discussed in Sec.~\ref{sec:hydrostatic}.}

A larger imbalance than that of Eq.~\eqref{eq:epsdelta} would decouple from the expansion and it would collapse and become electrically neutral. The maximum imbalance that is  able to survive after decoupling is $\epsilon_{q}^{coll}\sim 5\times 10^{-39}$, with a very mild dependence on redshift, dropping to $\sim 4\times 10^{-39}$ at $z=20$ due to the decreasing rate of expansion (see \cite{Lyttleton1959} for the case of fully coupled excess charges). 

Interestingly enough, this imbalance lies within the range required to produce seeds for the galactic dynamos~\cite{Davis1999}.

\subsection{Charged black holes}

A different mechanism comes from black holes, which may remove and displace charges from hot gas plasma, leading to a local charge imbalance that results in a non-zero net magnetic moment within a neutral galaxy.

Massive star remnants {in the form of} black holes of $\sim 10$ solar masses have independent dynamics from the plasma. The charge evolution of these black holes in a fully ionized plasma {was modeled in~\cite{Araya23,Padilla24}. Electrons, being the lightest charge carriers, have a thermal velocity about $30$ times faster than ions, suggesting that black holes initially acquire a negative charge. When a black hole first acquires a small negative electron charge $Q_{\rm ini}<0$, {the vacuum cross-section for proton accretion significantly increases}, but this does not necessarily apply to black holes in a protogalactic plasma because the individual, charged black hole affects ions only out to a {Debye length, $\ell_{\rm Debye}$, of the plasma}.  Since the {innermost stable circular orbit, $r_{\rm isco}$, of a black hole and $\ell_{\rm Debye}$ are $\sim 10^4$~m} for stellar remnant black holes in a typical galactic plasma, this favors again electron accretion due to their higher thermal velocity and a further increase in the negative sign charge for stellar mass black holes.

The black hole charge can continue growing until it reaches a {value,  $Q_{\rm pairs}$}, where electron-positron pairs are spontaneously created due to vacuum polarization \cite{Gibbons75}, which can also be understood as the charge that produces a chemical potential in the outer horizon of the black hole being equal to the rest mass of the electron 
\begin{equation}
    m_e c^2=k_C Q_{\rm pairs}e/r_{\rm isco}.
\end{equation}
This implies that once this charge is acquired, the black hole can emit electrons via athermal Hawking emission even for Hawking temperatures well below the electron mass \cite{Lehmann}, as is the case for stellar black holes.
Incidentally, for black holes formed by the collapse of a baryonic star, i.e. with a mass $M=\mu m_p N_{\rm baryons}$, {where $N_{\rm baryons}$ is the number of baryons in the black hole}, $Q_{\rm pairs}$ corresponds to a charge imbalance given by the ratio of Coulomb to Gravitational forces between an electron and a proton.  {This charge expected to be held by individual black holes is long-lived \cite{Padilla24} and it induces a galaxy wide charge, albeit with lower amplitude per unit mass.}

To see this, consider that in primeval galaxies, black holes formed in regions of cooled gas will eventually leave these and enter into contact with the plasma. If the black hole enters again a neutral gas region, its charge becomes stable due to the lack of free charges.  
This basically implies that a stellar mass black hole will accrete electrons up to the point where the Coulomb and Gravitational forces match, maintain the imbalance, and redistribute this charge within a galaxy.

We consider halo stars, so as to allow the black holes formed from them to be roughly velocity dispersion supported and to extract $Q_{\rm pairs}$ from the ionized gas in a few seconds. Assigning these black holes a fraction of $10^{-3}$ of the total mass of the galaxy, considering star formation efficiency as in \cite{Behroozi13}, leads to a {galaxy-wide charge imbalance} of,
\begin{equation}
    \epsilon_{\rm q}^{SBH}\sim 10^{-41},
\end{equation}
quite low but within the required amplitude.  This charge imbalance could in principle last long enough {to produce a seed field until} astrophysical dynamos start operating, before large portions of the plasma become neutral due to cooling.

Considering primordial black holes as dark matter candidates, stellar mass primordial black holes would also acquire a charge in a similar way. Current constraints \cite{sureda} suggest that their abundance could be as high as $10^{-2}$ of the galaxy mass, enabling a {charge imbalance and} seed field about a factor of $10$ larger than that of stellar black holes.

\subsection{Charge asymmetry in hydrostatic equilibrium}
\label{sec:hydrostatic}
It has been shown long ago~\cite{Pannekoek22,Rosseland24,Eddington1926} that due to the mass difference between electrons and the positive charged ions, a plasma in thermodynamic equilibrium in a gravitational field has a tiny positive charge which is basically determined by the relative strength between gravitational and Coulomb forces.  This is a global charge asymmetry, different from charge fluctuations that correspond to stochastic variations around the equilibrium value.
As we show below, this tiny charge excess is of the order of the square of the ratio of the Debye length to the virial radius of the system, which explains why this asymmetry is safely neglected in most applications.

For a protogalaxy in hydrostatic equilibrium, the gravitationally induced charge can be estimated taking a Boltzmann distribution for the number densities of the main plasma components, i.e.~electrons and protons, with a common and approximately constant temperature $T$, namely
\begin{eqnarray}
    n_e &=& n_e^0 \, e^{(-m_e \phi + e \psi)/k_B T}, \\
    n_p &=& n_p^0 \, e^{(-m_p \phi - e \psi)/k_B T},
\end{eqnarray}
where $k_B$ is the Boltzmann constant, $\phi=\phi(r)$ is the gravitational potential at a distance $r$ from the center, $\psi=\psi(r)$ is the (induced) electric potential, and $n_{e,p}^0$ are the number densities of electrons and protons, respectively, for zero gravitational and electric potentials (see e.g.~\cite{Landau:1980mil}). The quasi-neutrality condition, $n_e(r) \simeq n_p(r)$, therefore implies that $-m_e \phi + e \psi = -m_p \phi - e \psi$, which leads to
\begin{equation}
    \psi = - \frac{(m_p-m_e)}{2 e} \phi \simeq - \frac{m_p}{2 e} \phi.
\end{equation}
Using the Poisson equations for the electric potential, $\nabla^{2} \psi = - 4 \pi k_C \rho_{q}$ and the gravitational potential, $\nabla^2 \phi = 4 \pi G \rho_m$, the net charge density in the plasma can be related to the source of the gravitational field,
\begin{equation}
    \rho_{q} = \frac{G m_p}{2 k_C e}\, \rho_m \, .
\end{equation}
Note that this gives a charge imbalance with the same (positive) sign along the whole protogalaxy, i.e.~equilibrium implies an apparently tiny effect but which is coherent over galactic scales. As argued in \cite{Bally78} this conclusion actually applies to all self-gravitating systems larger in size than their Debye length (see also the derivation of the charge imbalance including the Debye length in e.g.~the chapter 3 of \cite{Parks2018})~\footnote{A side comment might be interesting at this point: In a very recent model of the solar corona \cite{barbieri24}, the electric field arising from the gravitationally induced charge asymmetry in a collisionless bound plasma has been included, and it was found to vanish in the stationary state. The key difference with the model we are considering comes from the border conditions: in the model of~\cite{barbieri24} the number of electrons and ions is kept fixed and constant within a semicircular tube, representing a coronal loop of confined plasma driven by strong magnetic fields (see for instance \cite{Reale2014}).  In our application magnetic fields are incipient.}. Also note that the charge asymmetry in equilibrium is very small and therefore the plasma can be considered as electrically neutral for most purposes. Using Eq.~\eqref{eq:rhoq}, the resulting imbalance is 
\begin{equation}\epsilon_q=\frac{G \mu m_p^2}{2 k_C e^2}\frac{\rho_m}{\rho_b},
\end{equation} 
which is remarkably similar to the gravitational to Coulomb force ratio of Eq.~\eqref{Eq:CG}, and significantly larger than the one in Eq.~\eqref{eq:epsdelta}, 
indicating that this imbalance is likely larger than leftover primordial ones and imbalances resulting from black hole charge displacement within galaxies. To give an idea of the amplitude in terms of physical properties of the plasma, by taking the plasma temperature equal to the virial temperature we can recast the last equation in terms of the Debye length (of the order of kilometers) and the virial radius of the system,  $\epsilon_q\approx(\ell_{\rm Debye}/R_{\rm vir})^2.$

\begin{figure}
    \centering
    \includegraphics[scale=.44]{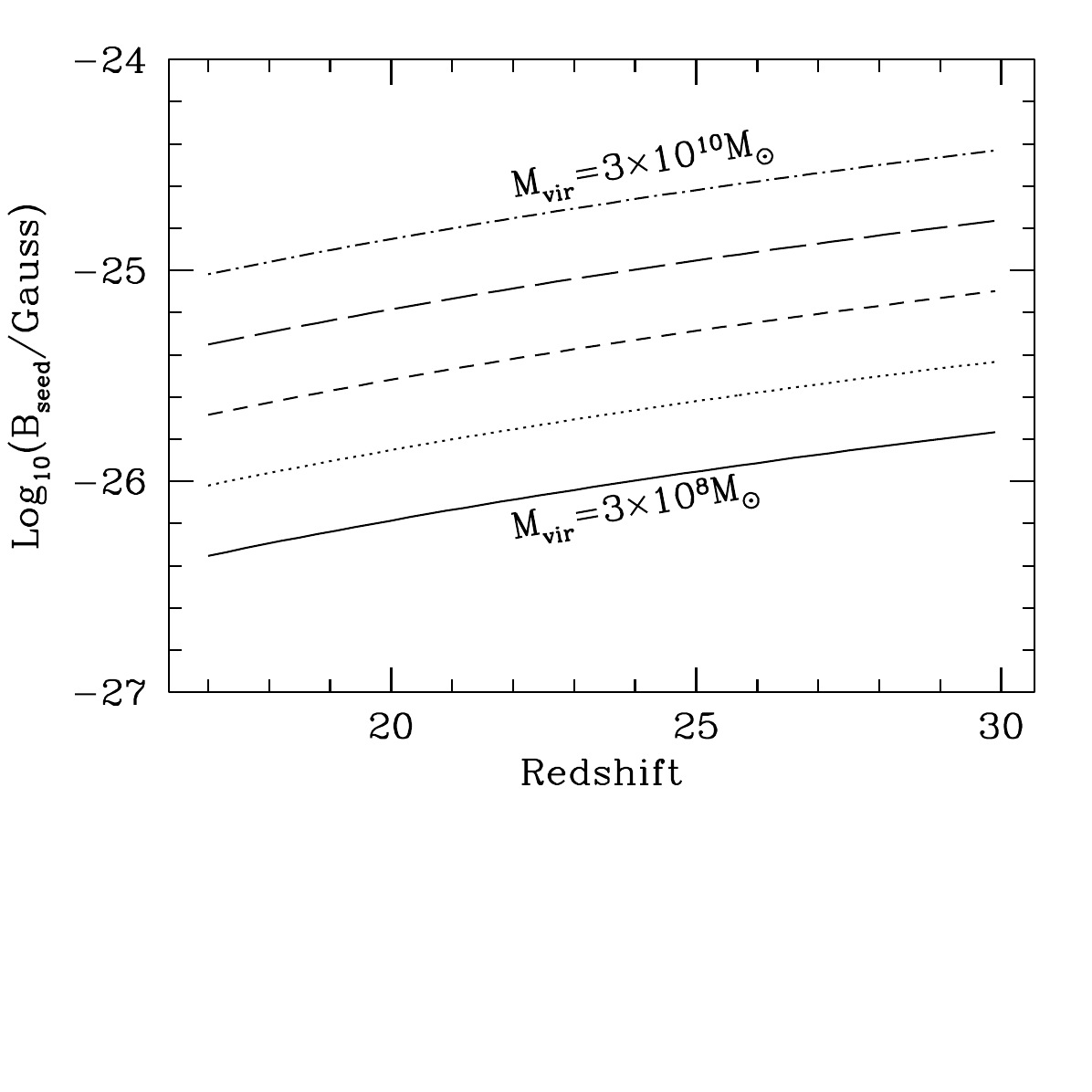}
    \vskip -2.5cm
    \caption{Dependence of the seed magnetic field on redshift for different choices of virial mass (different lines) going from $M_{\rm vir}\simeq3\times10^{10}M_\odot$, to $M_{\rm vir}\simeq3\times10^8M_\odot$ in half decade intervals.  A typical Milky Way progenitor at $z\sim20$ would lie on the short-dashed line.  The choice of this redshift has a mild impact on the seed field from charge imbalances in quasi-hydrostatic equilibrium.
    }\label{fig}
   \end{figure}
   
Considering that $\rho_m \simeq \frac{\Omega_m}{\Omega_b}\rho_b$ and replacing this relation in 
Eq.~\eqref{eq:B}, we obtain $B_{\rm seed} \simeq 3 \times 10^{-26}$~G, in the upper range of the minimum required seeds \cite{Davis1999}.  This seed arises from the different masses of electrons and ions, combined with {the initial rotation of a proto-galaxy inferred from galaxy formation models with a dominant dark-matter component.}
In Fig.~\ref{fig} we show the dependence of the amplitude of the seed magnetic field on redshift for different fixed values of halo virial mass.  {Notice the very mild dependence on redshift, which makes our estimate robust under the choice of collapse redshift for the protogalaxy}. 

\section{Discussion}
\label{sec:discussion}
As already summarized in the introduction, there are many proposals for cosmic magnetic field seeds. 
Several of them are based on astrophysical processes that most certainly must have occurred, although the magnitude of the seeds and the coherence length may not be trivial to determine. On general grounds and compared to the more primordial seeds, the magnitudes of these astrophysical seeds are typically smaller and require significant amplification by dynamo processes, but the coherence lengths are larger.

Some recent studies include~\cite{Hutschenreuter18}, where the evolved primordial field seeded by density perturbations during the radiation dominated epoch \cite{Matarrese2005,gopal2005,takahashi2005,Saga15} is found to be somewhat above $10^{-27}$~G in galaxy clusters. Another mechanism considering the free electrons left over after recombination and based on the growth of linear perturbations was presented in \cite{Naoz2013}, yielding magnetic seeds of order $10^{-25}$ G on comoving scales of $\sim$ ten kpc (see also~\cite{Flitter2023, Lee2024}). In~\cite{Attia21} the Biermann battery effect during the epoch of reionization was studied and three different Biermann battery channels were identified.

There are also new proposals for seeds generated by the first cosmic rays which can drive a Biermann battery in an inhomogeneous plasma, with estimated magnitudes of $\sim 5 \times 10^{-21}-10^{-23}$ G for one kiloparsec \cite{Ohira20, Ohira21} at $z \sim 20$ (with a strong decrease with the length scale and a strong dependence on the parameters for  early cosmic rays). 

In this work we have focused on the size and possible origins of tiny charge asymmetries that can generate adequate magnetic field seeds in slowly rotating protogalaxies.
Although there are possible charge asymmetries of primordial origin that cannot be ruled out, and black holes could induce these within galaxies, 
the charge imbalance present in the state of (quasi) hydrostatic equilibrium gives the largest seeds among the possibilities we consider, with magnetic field seeds of $\sim 10^{-25}$~G being generated, with a very tight and simple connection to the mass distribution.  This is a remarkably simple effect producing seeds in the upper range estimated in \cite{Davis1999}.  Moreover, it is interesting to note that, for instance, at $z \sim 20$ our estimates are of the same order as 
results from numerical simulations of Biermann battery processes like in \cite{Attia21} (as Fig.~\ref{fig} shows, our estimates, which cease to be of importance once astrophysical dynamos set in, do not depend strongly on redshift). Only at later times these simulations are able to provide  stronger magnetic fields at galactic scales  of up to $B \sim 10^{-20}$~G
by the end of reionization at $z=6$ including astrophysical, out of equilibrium processes such as supernovae winds \cite{Attia21,Durrive2015}. 
We also note that in~\cite{Widrow11} a simple estimate is given for the Biermann battery effect in a protogalaxy of $B_{\rm seed}\sim 10^{-19}$~G, although the vorticity is assumed to be the one present in the solar neighborhood today. 

An estimate that is similar to the third option studied here, considering the charge imbalance in hydrostatic equilibrium, was given previously in~\cite{Bally78}, quoting an amplitude $B \sim10^{-25}$~G without details on the calculation. However, our derivation is based on modern and detailed assumptions 
at the time that galaxy formation begins, when a seed is of interest, which requires the use of current understanding of dark matter dominated galaxies and their mass accretion history. 

Our scenario allows the presence of a magnetic field in the center of a rotating hot, ionized gas halo within which a galaxy disc starts to form.  Notice that we are not including vorticity on scales smaller than the galaxy, since these appear after gas dynamical processes start to affect the internal dynamics of the galaxy.  Our estimate corresponds to the upper limit of the magnetic field arising from the primeval galaxy rotation once ionization of the gas is complete after the initial gravitational collapse.  In this sense we are considering vorticity only on the scale of the whole halo.  This single-scale vorticity is guaranteed at least observationally, even if its origin is still under discussion. 

Note that the seed mechanism proposed here would in principle apply  within galaxies being there in place by the time star formation begins which is also the time when astrophysical winds start to take effect. The latter, among other out of equilibrium effects, could then provide a way to expel magnetic fields and seed the intergalactic space \cite{Bertone06}.

One more comment is in order. As explained e.g.~in \cite{Belmont2013}, it is to be expected that in the outer regions of a self-gravitating plasma in quasi-hydrostatic equilibrium, the charge imbalance is actually larger than the calculated assuming perfect equilibrium, which could yield larger magnetic fields, particularly when thermal electron motions are of the same order of magnitude as the escape velocity.  This condition is met by the protogalaxies we focus on here even at small radii.  This could lead to a larger seed by several orders of magnitude (we will explore this in the near future).

\section{Conclusions}
\label{sec:conclusions}
Summarizing,  we have analyzed the possibility for a tiny charge asymmetry in slowly rotating protogalaxies to be a source of magnetic field seeds. Among the charge asymmetry sources we have considered, the charge imbalance gravitationally induced in plasmas in hydrostatic equilibrium yields the strongest seeds, of order {$\sim10^{-25}$~G} on 
protogalactic scales.
This result is based on simple physics and on the current model of galaxy formation and evolution.

The magnetic seeds we have studied here add to other astrophysical processes producing seeds with varying strengths and scale dependences, which are   
an interesting option looking forward to the state of the art of observations such as those of \cite{Geach2023}, where highly ordered magnetic fields have been observed at $z \sim 2.6$. Such high amplitude fields could be reconciled with 
initially tiny -but almost unavoidable- magnetic seeds {like the ones analyzed in this work} by strong dynamos in early mergers. Further studies need to be done to understand if these fields can also seed intergalactic or intra-cluster media.

\section*{Acknowledgments}
    N.D.P. was supported by a RAICES, a RAICES-Federal, and PICT-2021-I-A-00700 grants from the Ministerio de Ciencia, Tecnología e Innovación, Argentina. J.R. thanks the support by a grant Consolidar-2023-2027 from the Secretaría de Ciencia y Tecnología (SeCyT) de la Universidad Nacional de Córdoba (UNC) and also support from the Consejo Nacional de Investigaciones Científicas y Técnicas (CONICET), Argentina. The work of I.J.A. is funded by ANID FONDECYT grants No.~11230419 and~1231133. F.A.S. thanks support by grants PIP 11220130100365CO,
PICT-2016-4174, PICT-2021-GRF-00719 and Consolidar-2018-2020, from CONICET, FONCyT (Argentina) and SECyT-UNC.  

\bibliographystyle{JHEP}
\bibliography{references}

\end{document}